\documentclass[10pt]{article}
\usepackage[utf8]{inputenc}
\usepackage{authblk}
\usepackage{fancyhdr}
\usepackage{extramarks}
\usepackage{amsmath}
\usepackage{amsthm}
\usepackage{amsfonts}
\usepackage{siunitx}
\usepackage{tikz}
\usepackage[plain]{algorithm}
\usepackage{algpseudocode}
\usepackage{multirow}
\usepackage{booktabs}
\usepackage{graphicx}
\usepackage{subfigure}
\usepackage[colorlinks,linkcolor=black,anchorcolor=black,citecolor=black,urlcolor=blue]{hyperref}
\usepackage{longtable}
 \usepackage{color}
\usepackage{amsmath,bm}
\usepackage{booktabs}
\usepackage{mathtools}
\usepackage{amssymb}
\usepackage{caption}
\usepackage{capt-of}
\usepackage{mciteplus}
\usepackage{cite}

\usepackage{mathrsfs}
\usepackage[title,titletoc,toc]{appendix}
\usepackage{parskip}
\usepackage{soul}
\usepackage{textcomp}
\usepackage[colaction]{multicol}
\usepackage[switch]{lineno}
\usepackage{lipsum}
\usepackage{etoolbox}
\usepackage{longtable}
\usepackage{array}
\usepackage{tablefootnote}
\usepackage{ragged2e}
\newcolumntype{C}[1]{>{\centering\arraybackslash}p{#1}}
\usetikzlibrary{automata,positioning}
\usepackage{adjustbox}
\usepackage{makecell}
\usepackage{pdflscape}
\usepackage{setspace}
\usepackage{multicol}
\usepackage[utf8]{inputenc}
\usepackage[T1]{fontenc}
\usepackage{textgreek}
\usepackage{xcolor}
\usepackage{url}

\usepackage{xr}
\usepackage{hyperref}
\usepackage{xr-hyper}

\usetikzlibrary{automata,positioning}

\makeatletter
\newcommand*{\addFileDependency}[1]{
  \typeout{(#1)}
  \@addtofilelist{#1}
  \IfFileExists{#1}{}{\typeout{No file #1.}}
}
\makeatother

\begin{document}

\title{Proteomic Learning  of Gamma-Aminobutyric Acid (GABA) Receptor-Mediated Anesthesia} 

\author[1,2,*]{Jian Jiang}
\author[1]{Long Chen}
\author[1]{Yueying Zhu}
\author[1]{Yazhou Shi}
\author[1]{Huahai Qiu}
\author[1]{Bengong Zhang}
\author[3]{Tianshou Zhou}
\author[2,4,5,+]{Guo-Wei Wei}

\affil[1]{Research Center of Nonlinear Science, School of Mathematical and Physical Sciences, Wuhan Textile University, Wuhan, 430200, P R. China}
\affil[2]{Department of Mathematics, Michigan State University, East Lansing, Michigan 48824, USA}
\affil[3]{Key Laboratory of Computational Mathematics, Guangdong Province, and School of Mathematics, Sun Yat-sen University, Guangzhou, 510006, P R. China}
\affil[4]{Department of Electrical and Computer Engineering Michigan State University, East Lansing, Michigan 48824, USA}
\affil[5]{Department of Biochemistry and Molecular Biology Michigan State University, East Lansing, Michigan 48824, USA}
\affil[*]{Corresponding author: jjiang@wtu.edu.cn}
\affil[+]{Corresponding author: weig@msu.edu}

\date{\today} 

\maketitle

\abstract{

Anesthetics are crucial in surgical procedures and therapeutic interventions, but they come with side effects and varying levels of effectiveness, calling for novel anesthetic agents that offer more precise and controllable effects.  
Targeting Gamma-aminobutyric acid (GABA) receptors, the primary inhibitory receptors in the central nervous system,
could enhance their inhibitory action, potentially reducing side effects while improving the potency of anesthetics. 
In this study, we introduce a proteomic learning of GABA receptor-mediated anesthesia based on 24 GABA receptor subtypes by considering over 4000 proteins in protein-protein interaction (PPI) networks and over 1.5 millions known binding compounds. We develop a corresponding drug-target interaction network to identify potential lead compounds for novel anesthetic design. To ensure robust proteomic learning predictions, we curated a dataset comprising 136 targets from a pool of 980 targets within the PPI networks. We employed three machine learning algorithms, integrating advanced natural language processing (NLP) models such as pretrained transformer and autoencoder embeddings. Through a comprehensive screening process, we evaluated the side effects and repurposing potential of over 180,000 drug candidates targeting the GABRA5 receptor. Additionally, we assessed the ADMET (absorption, distribution, metabolism, excretion, and toxicity) properties of these candidates to identify those with near-optimal characteristics. This approach also involved optimizing the structures of existing anesthetics. 
Our work presents an innovative strategy for the development of new anesthetic drugs, optimization of anesthetic use, and  deeper understanding of potential anesthesia-related side effects.
} 


\maketitle

 {\setcounter{tocdepth}{4} \tableofcontents}
\newpage

\section{Introduction}
With the continuous advancement of medical technology, the critical role of anesthetics in surgery and treatment has become increasingly prominent. However, current anesthetics still have significant limitations and side effects, which urgently necessitates the development of new anesthetic agents in the medical research field \cite{vlisides2012neurotoxicity,zhang2023emerging}. Traditional anesthetics may lead to various side effects, such as respiratory depression, cardiovascular issues, and postoperative cognitive dysfunction, which pose potential risks to patient safety and postoperative recovery \cite{kotekar2018postoperative}. Additionally, there is variability in patients’ responses to anesthetics, making it challenging to achieve individualized anesthetic regimens. This necessitates the development of new anesthetic agents that provide more precise and controllable anesthetic effects \cite{zeng2024personalized}. Therefore, the development of new anesthetic agents with higher safety, fewer side effects, and greater individual adaptability can not only enhance the success rate of surgeries and treatments but also significantly improve the overall prognosis and quality of life for patients \cite{swain2017adjuvants,song2023necessity}. 

In the human central nervous system, GABA ($\gamma$-aminobutyric acid) receptors are the primary inhibitory receptors, maintaining neural system balance by regulating neuronal excitability \cite{karunarathne2023gamma}. As a major inhibitory neurotransmitter, GABA acts on these receptors to reduce neuronal activity, thereby inhibiting the transmission of neural signals \cite{koh2023gaba}.

GABA receptors are divided into two main categories: GABA\textsubscript{A} and GABA\textsubscript{B} \cite{qian2023current}. GABA\textsubscript{A} receptors are ligand-gated ion channels; when GABA binds to them, the receptor undergoes a conformational change, leading to the opening of the chloride ion channel, an influx of chloride ions into the cell, and subsequent hyperpolarization of the cell membrane, thereby inhibiting neuronal excitability \cite{karlsson2024ligand}. In contrast, GABA\textsubscript{B} receptors are G protein-coupled receptors that indirectly regulate potassium and calcium ion channels through the activation of secondary messenger systems \cite{fritzius2024preassembly}.

In the field of anesthesiology, the association between GABA receptors and anesthetics is particularly crucial \cite{muheyati2024extrasynaptic}. Many general anesthetics and sedatives exert their effects by enhancing the function of GABA\textsubscript{A} receptors. For instance, benzodiazepines (e.g., diazepam), barbiturates (e.g., thiopental sodium), and volatile anesthetics (e.g., isoflurane) increase the affinity of GABA for GABA\textsubscript{A} receptors or prolong the opening time of the chloride ion channel, thereby enhancing inhibitory neurotransmission and inducing sedation, anxiolysis, anticonvulsant, and anesthetic effects \cite{takano2024stereochemical,hassan2024sclareol}. Additionally, allosteric modulators of GABA receptors (e.g., propofol) are widely used in clinical anesthesia; by binding to different sites on GABA\textsubscript{A} receptors, they further enhance the action of GABA, producing potent anesthetic effects \cite{yang2024comparison}.

Protemic technology has increasingly presented a vast potential in anesthesia \cite{crowder2024systematized}, and the use of proteomic tools to study anesthetic binding sites has offered a better understanding of the mechanisms of anesthetic action. Protein-protein interaction (PPI) networks at the proteomics level provide a systematic framework for exploring potential therapeutic strategies and their possible side effects \cite{gao2021proteome}. These networks encompass the direct and indirect connections among various proteins, collectively driving the complex biological activities within an organism \cite{du2024multiscale}. Utilizing the powerful String v11 database (\href{https://string-db.org/}{https://string-db.org/}), we can obtain extensive and diverse PPI datasets related to specific proteins or diseases \cite{szklarczyk2019string,du2024multiscale}. In the field of anesthesia research, the PPI network associated with GABA receptors can be extracted using the String v11 database. Through in-depth analysis of this network, we not only gain insights into how drugs interact with GABA receptors but also potentially uncover new drug targets. This approach could optimize pharmacotherapy regimens and reduce the incidence of side effects.

Nevertheless, traditional methods for developing anesthetics often involve long duration, high cost, and limited screening efficiency \cite{lopes2024artificial}. To address these challenges, the application of machine learning (ML) techniques in drug development is increasingly gaining prominence \cite{ballester2010machine,feng2023machine,zhu2023tidal}. ML approaches have outperformed other competing methods in D3R Grand Challenges, an annual worldwide competition series in computer-aided drug design \cite{nguyen2019mathematical,nguyen2020mathdl}.
ML technologies process and analyze vast amounts of biomedical data, thereby enhancing the efficiency of drug design and screening \cite{shen2023svsbi}, predicting the biological activity and pharmacological properties of new molecules, optimizing drug structures, and improving binding specificity to GABA receptors. Moreover, they can predict the potential toxicity and side effects of drugs, enabling the screening of safer candidate drugs. Since ML approaches have discovered two primary evolutionary mechanisms of SARS-CoV-2 \cite{chen2020mutations,wang2021mechanisms}, they aid in discovering new mechanisms of action and drug targets, providing robust support for the development of innovative anesthetics.

In this study, we constructed a proteomics-based ML system aimed at exploring novel anesthetic drugs targeting GABA receptors. Utilizing the String v11 database, we extracted the protein interaction networks of 24 GABA receptor subtypes, considering these related proteins as potential therapeutic targets and sites that may induce side effects. We then collected experimental binding affinity (BA) data for these target proteins from the ChEMBL database and developed ML models based on this information. Compound information was transformed into two different latent vector fingerprints through transformer networks and autoencoder models. These molecular fingerprints, combined with support vector machines, formed our BA prediction model.
By cross-predicting over 180,000 compounds, we assessed their potential for side effects and reuse value. Using these models, we screened for promising lead compounds and conducted an in-depth analysis of side effects for FDA-approved drugs and other existing medications. Additionally, we optimized the molecular structures of existing drugs to reduce side effects and improve their pharmacokinetic properties. During the compound screening process, we also comprehensively considered pharmacokinetic parameters, namely absorption, distribution, metabolism, excretion, and toxicity (ADMET), as well as synthetic feasibility. Our platform is expected to facilitate the development process of anesthetic drugs.

\section{Results}
\subsection{The GABA receptors PPI networks}
GABA is the primary inhibitory neurotransmitter in the mammalian central nervous system, playing a crucial role in reducing neuronal excitability throughout the nervous system. GABA exerts its effects by binding to specific receptors, known as GABA receptors, which are divided into two main types: GABA\textsubscript{A} and GABA\textsubscript{B}.
GABA\textsubscript{A} receptors are ionotropic receptors that function as ligand-gated chloride channels. When GABA binds to these receptors, the channel opens, allowing chloride ions to enter the neuron. This influx of chloride ions causes hyperpolarization of the neuronal membrane, making it less likely for the neuron to fire an action potential. GABA\textsubscript{A} receptors act quickly and are primarily responsible for the rapid inhibitory effects of GABA.
In contrast, GABA\textsubscript{B} receptors are metabotropic receptors that are coupled to G-proteins. Activation of GABA\textsubscript{B} receptors initiates a signaling cascade that can lead to the opening of potassium channels and inhibition of adenylate cyclase activity. This results in the slow and prolonged inhibitory effects of GABA, contributing to the overall inhibitory tone in the nervous system.

The subunits of GABA\textsubscript{A} receptors include $\alpha$ (GABRA1, GABRA2, GABRA3, GABRA4, GABRA5, GABRA6), $\beta$ (GABRB1, GABRB2, GABRB3), $\gamma$ (GABRG1, GABRG2, GABRG3), $\delta$ (GABRD), $\epsilon$ (GABRE), $\pi$ (GABRP), $\theta$ (GABRQ), and $\rho$ (GABRR1, GABRR2, GABRR3). In contrast, the GABA\textsubscript{B} receptor has two subunits: GABBR1 and GABBR2. Additionally, we collected other genes related to GABA receptors, including GABARAP, GABARAPL1, and GABARAPL2. In total, there are 24 targets associated with GABA receptors.
We extracted the corresponding 24 PPI networks by sequentially entering these gene names into the STRING database. Within each network, there is a core sub-network of proteins that interact directly with the GABA receptor, while the directly and indirectly interacting proteins together form the global network. We limited the number of proteins in each global network to 201. These 24 networks with total 4824 proteins are not completely independent, as some overlapping proteins are present. After removing the overlap proteins, 980 proteins are left in 24 PPI networks.

Compounds that act as agonists or antagonists of the GABA receptor exhibit pharmacological effects in anesthesia, which encourages the search for additional compounds that bind to the GABA receptor. The desired drugs must demonstrate specificity for the target protein without causing adverse side effects on other proteins. To evaluate the binding effects of small molecules on receptor proteins and other proteins within the PPI network, we collected the SMILES strings for each protein from the ChEMBL database and developed ML models. These models were then used to systematically analyze the side effects and repurposing potential of inhibitor compounds.
We gathered a total of 136 datasets, which included sufficient inhibitor data points for 980 proteins within the 24 extracted PPI networks, encompassing a total of 183,250 inhibitor compounds. Additionally, we compiled an inhibitor dataset for the human ether-\`{a}-go-go-related gene (hERG) protein and developed appropriate ML models \cite{feng2023virtual} (see  Supporting Information S6). 

\begin{figure}[htp]
    \centering
    \includegraphics[width=15cm]{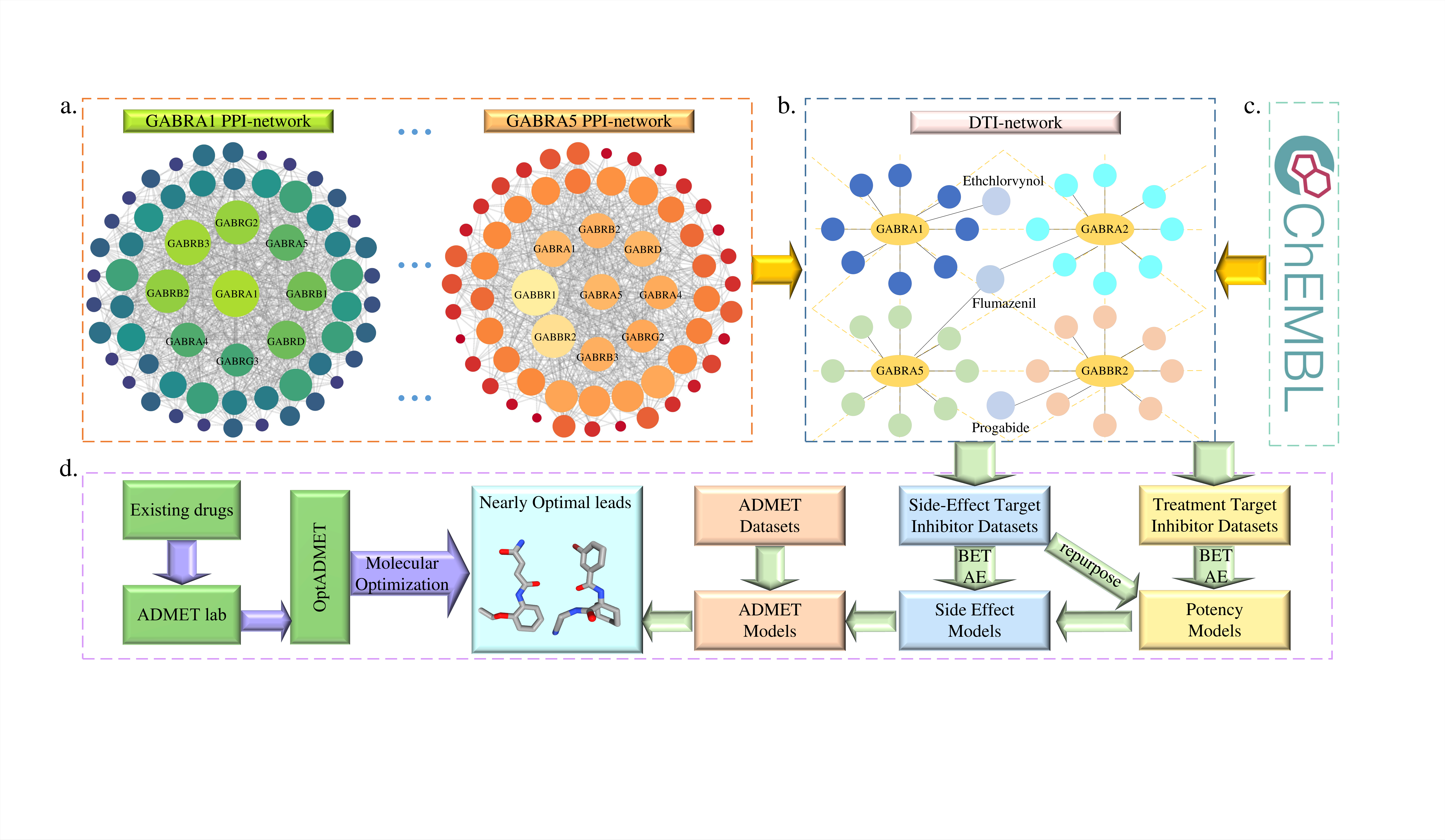}
        \caption{Flowchart of nearly optimal lead compounds screening for Gamma-aminobutyric acid (GABA) receptor agonists. a: The protein-protein interaction (PPI) networks of 24 GABA receptor subtypes involve 4824 proteins, and each receptor subtype has a core and global PPI network. Here only two PPI networks (GABRA1 and GABRA5) with several compounds are shown for simplicity. For more detailed information on the PPI networks, please refer to Table of the Supporting Information. b: The drug target interaction (DTI) network constructed against GABA receptors include 136 targets and 183250 inhibitor compounds that are collected from ChEMBL database in c. Here, only four targets (GABRA1, GABRA2, GABRA5, and GABBR2) with a few compounds are presented for simplicity. The yellow dashed lines mean the connections among 136 targets. d: Nearly optimal lead compounds were screened by two technical routes, the first being a predictive model for side effects and repurposing assessment as well as an ADMET screening model, and the second being molecular optimization of existing drugs.}
        \label{fig1}
\end{figure}

The framework of this study is illustrated in Figure \ref{fig1}. For the 24 anesthesia-related GABA receptors, we employed a proteomics-based approach to identify potential side effect targets using PPI networks. As shown in Figure \ref{fig1}a, we identified 980 unique targets from 4,824 proteins within the 24 PPI networks. Due to the lack of corresponding datasets for some targets and the insufficient data volume for others, we excluded these targets and ultimately obtained 136 targets from 980 ones. Specifically, the inhibitor compound should be homo sapiens and single protein, and the minimal training number should be larger than 250 to ensure the reliable ML prediction. Among these 136 targets, one was the therapeutic target GABRA5, while the remaining 135 targets were designated as side effect targets (including hERG as one of the side effect targets).
Simultaneously, we collected several common anesthetics that interact with GABA receptors and constructed a drug-target interaction (DTI) network, as depicted in Figure \ref{fig1}b. The drug compounds datasets shown in Figure \ref{fig1}c are collected from ChEMBAL database (https://www.ebi.ac.uk/chembl/). Utilizing the constructed PPI and DTI networks, we designed two technical routes to identify and generate nearly optimal lead compounds, as demonstrated in Figure \ref{fig1}d. The first route begins with the PPI network and applies prediction models for side effect and repurposing evaluations, as well as ADMET screening models. The second route starts with the DTI network, where existing drugs undergo ADMET screening; for molecules that do not meet ADMET criteria, molecular optimization is performed using the OptADMET online server\cite{yi2024optadmet}. Through these steps, we aim to identify or generate anesthetics with excellent properties.

\subsection{Binding affinity predictions}
Figure S1 in the Supporting Information presents a heatmap illustrating cross-target binding affinity (BA) predictions using 136 machine learning (ML) models. The diagonal elements represent the Pearson correlation coefficient ($R$) obtained from the ten-fold cross-validation of our models. Among the 136 models, two achieved $R$ values greater than 0.9, while 80 models had $R$ values exceeding 0.8. The lowest $R$ value of 0.497 was obtained from the model built using the CALCR inhibitor dataset, and the average $R$ value across all models is 0.791. Furthermore, the root-mean-square deviation (RMSD) values of these models fall within a reasonable range of [0.468, 1.406] kcal/mol, with an average RMSD value of 0.880 kcal/mol, as detailed in Table S1 of the Supporting Information. Based on our calculated $R$ values and RMSD values, these models demonstrate extremely high predictive accuracy and are reliable for BA prediction.

\subsubsection{Cross-target binding affinity predictions}
Cross-target binding affinity (BA) predictions are a powerful tool in various aspects of drug development. By leveraging machine learning   models, molecular docking, and network analysis, it is possible to effectively evaluate the multi-target activities of compounds, aiding in the identification of potential side effects, drug repurposing opportunities, and the design of multi-target drugs. In Figure S1, the non-diagonal elements represent the highest BA values (with the maximum absolute value) predicted by various models for inhibitor compounds within a dataset. The identifiers on the left side of the heatmap denote the 136 inhibitor datasets, while the symbols at the top correspond to the 136 ML models. Therefore, each column illustrates the predictions of a specific model. Specifically, the $i$th element in the $j$th column represents the prediction of the $j$th model for the $i$th dataset. These cross-target predictions reveal the potential side effects of one inhibitor dataset on other proteins. According to the literature, a widely accepted inhibition threshold is a BA value of -9.54 kcal/mol (K$_i$ = 0.1 $\mu$M, referring to the inhibition constant) \cite{flower2002drug}. At this threshold, out of 18,496 cross-predictions, 15,318 were found to indicate side effects, as the predicted maximum BA was less than -9.54 kcal/mol. Conversely, the remaining 3,178 cross-predictions, with maximum BA values greater than -9.54 kcal/mol, suggest weaker side effects.

The color of the non-diagonal elements represents the intensity of the side effects, with lighter colors indicating weaker effects. From all cross-target predictions shown in Figure S1, several light vertical lines can be observed, indicating very mild predicted side effects for these proteins. This phenomenon can largely be attributed to the label distribution, where the majority of collected experimental binding affinities (BAs) exceed -9.54 kcal/mol. In such instances, the predictive capabilities of the machine learning models may be limited. Off-target effects, where a drug binds to proteins other than its intended target, can lead to unintended and potentially harmful side effects. Therefore, identifying similar binding sites on off-target proteins is crucial in drug design to predict and mitigate these adverse effects. Proteins within the same family often exhibit similar three-dimensional (3D) structures or protein sequences, resulting in analogous binding sites. Inhibitory compounds effective against one protein may bind to other proteins within the same family. Figure S1 reveals five targets—SSTR1, SSTR2, SSTR3, SSTR4, and SSTR5—that exhibit potential for mutual side effects. These proteins belong to the somatostatin receptor family, which specifically mediates the effects of the peptide hormone somatostatin. As illustrated in Figure S2 of the Supporting Information, these receptors share similar 3D conformations and 2D sequences.

Furthermore, additional examples of mutual side effects can be observed within other protein families. For instance, muscarinic acetylcholine receptors (CHRM1, CHRM2, CHRM3, and CHRM4), dopamine receptors (DRD1, DRD2, DRD3, DRD4, and DRD5), prostaglandin E receptors (PTGER1, PTGER2, PTGER3, and PTGER4), and sphingosine-1-phosphate receptors (S1PR1, S1PR2, S1PR3, and S1PR4) all demonstrate the potential for mutual side effects. These instances underscore the structural and functional similarities within protein families, indicating that drugs targeting one specific protein may also interact with other family members, leading to unintended physiological responses and side effects. Therefore, understanding these interactions is crucial for the development of highly selective and specific therapeutic agents, which can minimize off-target effects and improve overall drug efficacy and safety. By considering these potential mutual side effects, researchers can better predict and mitigate adverse effects, ultimately advancing the design of targeted therapies with higher precision and reduced side effects.

\subsubsection{Predictions of side effects and repurposing potentials}
In the process of drug development, cross-target prediction serves as an essential tool for effectively detecting potential side effects and assessing the repurposing potential of inhibitors. Side effects typically arise when a candidate drug shows strong BA to the target protein while unintentionally acting as a potent inhibitor for other proteins. Therefore, predicting these side effects is crucial for ensuring the safety of the drug. Conversely, if a candidate drug demonstrates weak BA to the target protein but presents strong inhibitory effects on other proteins, it is considered to have repurposing potential. Drug repurposing, which involves exploring new therapeutic indications for existing drugs, not only significantly reduces research and development costs but also accelerates the drug approval process \cite{chen2024machine}.

\begin{figure}[htp]
    \centering
    \includegraphics[width=15cm]{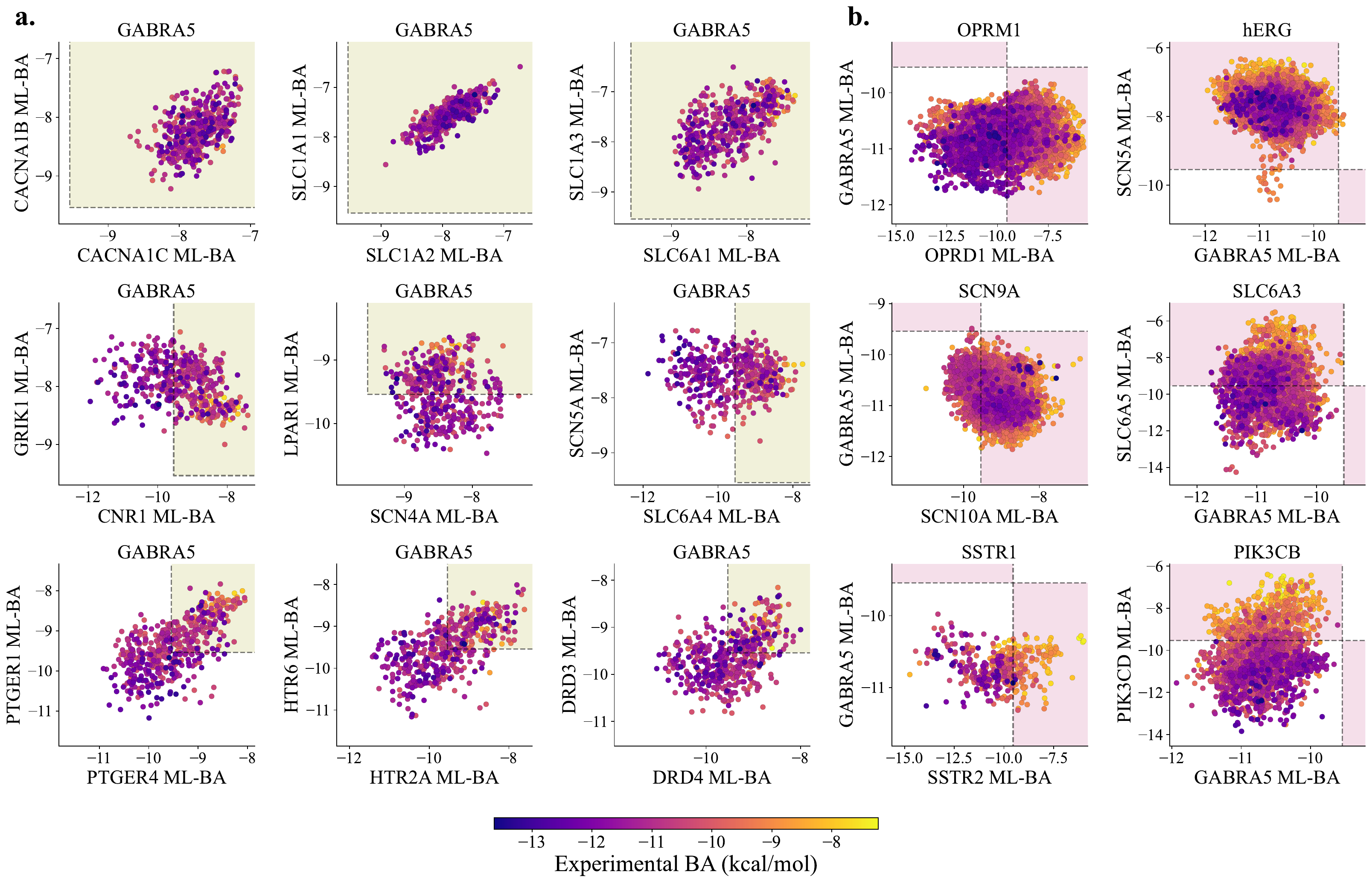}
        \caption{Examples of side effect and repurposing potential prediction. a: Three rows of inhibitors targeting GABRA5 are presented, each causing side effects on zero, one, and both of the two side effect targets, respectively. The yellow frames indicate no side effects. b: Inhibitors of GABRA5 with repurposing potential are shown, where the pink frames indicate an inhibitor's repurposing potential for one therapeutic target without side effects on the other.}
        \label{fig3}
\end{figure}

For instance, Figures \ref{fig3}a and \ref{fig3}b illustrate specific cases of side effects and repurposing, respectively. Each subplot depicts a target protein along with its two off-target proteins or side effect target proteins. Specifically, the title of each panel, as well as the $x$- and $y$-axes, represent the target protein and two distinct off-target proteins, respectively. Furthermore, the color of the scatter points reflects the experimental BA values of the inhibitors for these proteins, with dark blue indicating high affinity and yellow indicating low affinity. Additionally, the $x$- and $y$-axes, respectively, display the predicted BA values obtained from two ML models constructed using inhibitor datasets for the two off-target proteins.

As illustrated in Figure \ref{fig3}a, the yellow frames in nine panels highlight regions predicted to have no side effects on two specific off-target proteins. The three rows in Figure \ref{fig3}a display examples of inhibitors targeting a designated protein with side effects on zero, one, and two of the given off-target proteins, respectively. In the subplot located in the first row and first column, all active inhibitors targeting the therapeutic target GABRA5 are predicted to have no side effects on the two targets, CACNA1C and CACNA1B, as their BA values are both greater than -9.54 kcal/mol. In the subplot situated in the second row and second column, all inhibitors of GABRA5 show no side effects on target SCN4A, while approximately half of the GABRA5 inhibitors exhibit side effects on target LPAR1. Furthermore, in the subplot found in the third row and third column, most active inhibitors of GABRA5 demonstrate side effects on both DRD3 and DRD4 proteins.

In addition to assessing side effects, our cross-target BA predictions can also indicate potential for drug repurposing. Each subplot in Figure \ref{fig3}b highlights instances where some inactive inhibitors for off-target proteins are predicted to be potent inhibitors for other proteins, as indicated by the pink frames. Specifically, certain inhibitors for the off-target protein exhibit very weak BA values (i.e., predicted BA greater than -9.54 kcal/mol), but they demonstrate strong BA values for the therapeutic target GABRA5 (i.e., predicted BA less than -9.54 kcal/mol). Notably, in the subplot located in the first row and first column of Figure \ref{fig3}b, many inactive inhibitors of OPRM1 are identified as having repurposing potential towards GABRA5 while showing no repurposing potential for the off-target OPRD1. It is important to note that the OPRD1 and OPRM1 genes encode the δ-opioid receptor and μ-opioid receptor, respectively, both of which are involved in modulating pain perception and analgesia, making them targets for analgesic drugs. Furthermore, we observed that in cases where these inactive inhibitors exhibit repurposing potential for GABRA5, they tend to have minimal side effects on other off-target proteins.

\subsubsection{Protein similarity inferred by cross-target BA correlations}
\begin{figure}[htp]
    \centering
    \includegraphics[width=15cm]{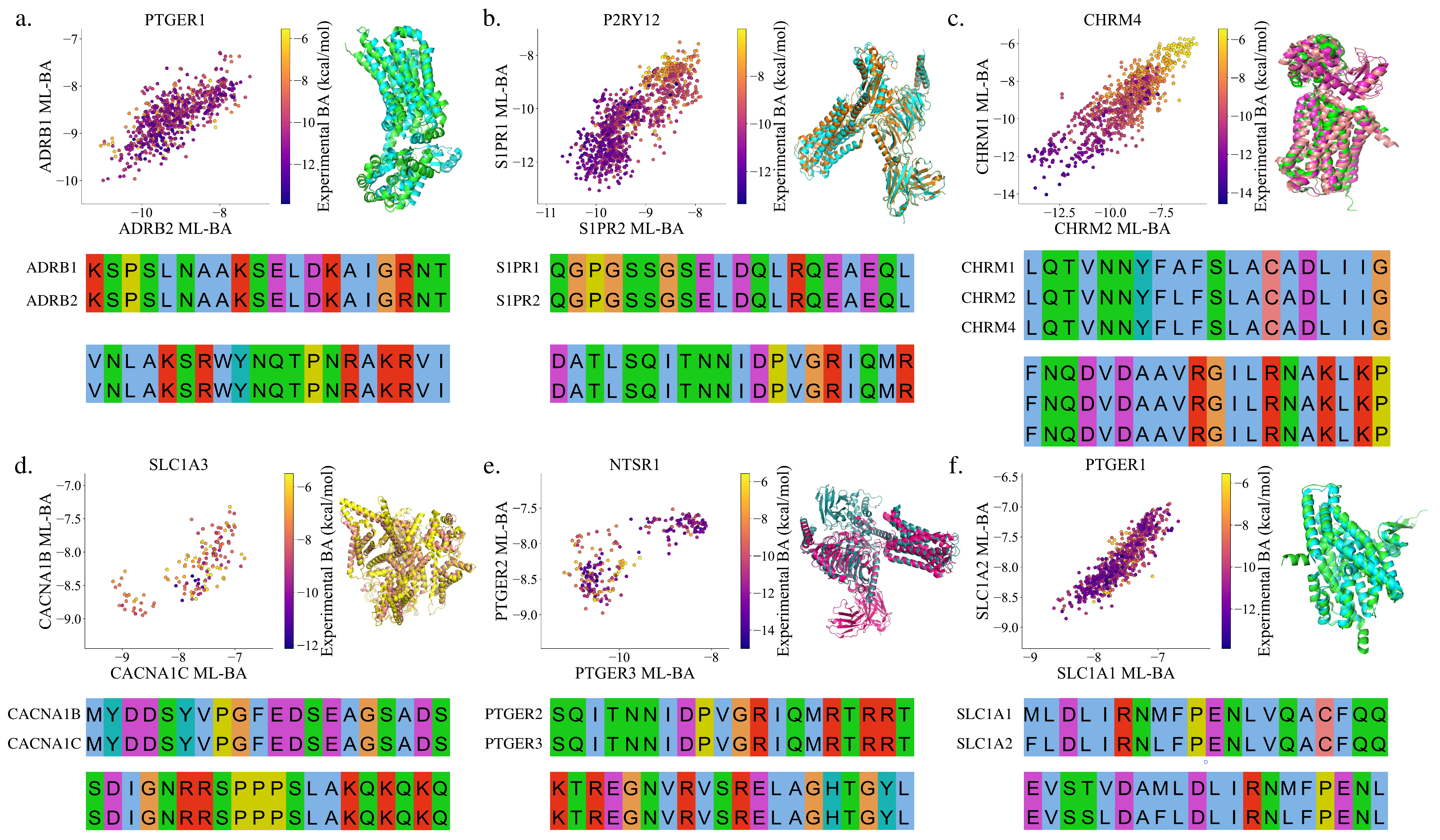}
        \caption{Six examples of related predicted BAs illustrate the sequence and structural similarities of proteins. In each example, the $x$ and $y$ axis of the panel display the predicted BA values for two other proteins. On the right side of the scatter plot, the 3D structural alignment is shown, while the 2D sequence alignment is displayed below. The 3D structures used for alignment include PDB 7BU7 and 4LDE for ADRB1 and ADRB2 (a), PDB 7VIE and 7T6B for S1PR1 and S1PR2 (b), PDB 6ZFZ, 3UON, and 5DSG for CHRM1, CHRM2, and CHRM4 (c), PDB 7MIX and 8WE6 for CACNA1B and CACNA1C (d), PDB 5U09 and 5ZTY for CNR1 and CNR2 (e), and PDB 7CX2 and 7WU9 for PTGER2 and PTGER3 (f).}
        \label{fig4}
\end{figure}

Binding site similarity can generate cross-target BA correlations. Conversely, high BA correlation can help identify binding site similarities. According to our cross-target predictions, there are several examples of high BA correlation associated with similar binding sites. For instance, as shown in Figure \ref{fig4}a, the predicted BA of PTGER1 inhibitors for ADRB1 and ADRB2 proteins exhibits an almost linear correlation, with a Pearson correlation coefficient ($R$) of 0.730. This high correlation is attributed to binding site similarity, which is validated through 3D protein structure and 2D sequence alignment, as depicted in Figure \ref{fig4}a. We found that the 3D structures of ADRB1 and ADRB2 proteins are highly similar, and the 2D sequence identity near the binding sites is approximately 77.49\%. Specifically, in Figure \ref{fig4}b, the predicted BA of P2RY12 inhibitors for S1PR1 and S1PR2 proteins shows an $R$ value of 0.798. Notably, the 2D sequence identity near their binding sites is 100\%. Both S1PR1 and S1PR2 are receptors for sphingosine-1-phosphate (S1P) and are involved in regulating cell migration, vascular integrity, and immune cell function. Additionally, in Figure \ref{fig4}c, a bilinear correlation between predicted and experimental BA is observed because the target protein and the two off-target proteins belong to the same protein family. Specifically, the predicted BA of CHRM4 inhibitors for CHRM1 and CHRM2 proteins shows an $R$ value of 0.851, with a 2D sequence identity near the binding sites of 75.56\% between CHRM1 and CHRM2. According to our further calculations, the 2D sequence identity near the binding sites is 76.40\% between CHRM1 and CHRM4, and 84.30\% between CHRM2 and CHRM4. These results suggest that potent CHRM4 inhibitors are likely to be strong binders for both CHRM1 and CHRM2 proteins as well. Figure \ref{fig4}d-f present similar binding sites between CACNA1B and CACNA1C, CNR1 and CNR2, PTGER2 and PTGER3, respectively. 

\subsubsection{Repurposing to GABA receptors and side effect on hERG}
GABRA5, as a critical neurotransmitter receptor subtype, plays a significant role in the pharmacological research of anesthetics \cite{zurek2012inhibition,shan2017mirnas,feng2023virtual}. To evaluate the BA of anesthetics for the GABRA5 receptor, we employed a multi-target prediction strategy to screen anesthetics with potential for repurposing. Our collected dataset comprises 136 datasets encompassing 183,250 compounds, providing a rich source of candidates for investigating potential drugs targeting the GABRA5 receptor. 

Considering the risk of hERG-related side effects, we utilized ML models to predict the BAs of these anesthetics to other crucial targets. Specifically, we set a very stringent side effect threshold of -8.18 kcal/mol (K$_i$=1 $\mu$M) to ensure the safety of candidate drugs in clinical applications. If the predicted BA exceeds the set threshold, the anesthetic is considered to have a lower risk of side effects, thus qualifying it as a preferred drug for further study. Figures S10, S11, S12, and S13 illustrate the predicted BAs of the remaining 134 datasets to both GABRA5 and hERG. The yellow frames highlight regions where compounds may exhibit potential for repurposing as GABRA5 modulators without eliciting hERG side effects. As observed in these figures, we found that all datasets contain a substantial number of compounds within these yellow-highlighted regions.

\subsection{Druggable property screening}
The ADMET properties (Absorption, Distribution, Metabolism, Excretion, and Toxicity) are crucial determinants of a drug's pharmacokinetics and safety profile \cite{m2013pharmacokinetic,saravanan2024deep}. Absorption and distribution influence bioavailability and tissue targeting, while metabolism and excretion determine the drug's duration of action and clearance. Toxicity assessment is essential for ensuring the drug's safety at therapeutic doses. Early optimization of these properties can reduce late-stage failures and enhance the likelihood of clinical success. Additionally, the hERG channel plays a critical role in evaluating the cardiac safety of new drugs, as its inhibition can lead to life-threatening arrhythmias \cite{feng2023virtual}. Assessing hERG activity early in the drug development process helps identify and mitigate potential cardiotoxicity risks. Prioritizing hERG screening can reduce the likelihood of late-stage clinical failures and ensure safer therapeutic profiles.

In this section, to identify potential anesthetics, we need to systematically screen the ADMET properties, synthetic accessibility (SAS), and hERG risk of all datasets. Specifically, we focus on six ADMET parameters: FDAMDD, T\textsubscript{1/2}, F\textsubscript{20\%}, log P, log S, and Caco-2. More specifically, FDAMDD (FDA Maximum Daily Dose) indicates the maximum daily dose of a drug as stipulated by the U.S. Food and Drug Administration, ensuring the drug is used within a safe dosage range \cite{li2024crossfuse}. T\textsubscript{1/2} (half-life) represents the time required for the drug concentration in the body to decrease by half, reflecting the drug's clearance rate and duration of action \cite{abdelgadir2024determine}. F\textsubscript{20\%} (bioavailability) denotes the proportion of the drug that enters systemic circulation and exerts its therapeutic effect; for instance, if it is 20\%, it means that 20\% of the drug is available in the bloodstream to produce its effect \cite{tian2011adme}. Additionally, log P (partition coefficient) reflects the drug's distribution balance between lipid and aqueous phases (e.g., octanol/water), which is crucial for evaluating the drug's lipophilicity and hydrophilicity, influencing its absorption and distribution \cite{bodor1989new}. Log S (aqueous solubility) indicates the drug's solubility in water, expressed logarithmically, directly affecting its absorption and bioavailability in the body \cite{win2023using}. The Caco-2 model, utilizing human colon adenocarcinoma cells, simulates the drug's permeability through intestinal epithelial cells, serving as an essential \textit{in vitro} model for assessing oral drug absorption \cite{sun2008caco}. Furthermore, SAS is used to evaluate the case of synthesizing a compound, which is significant for drug development and optimization.

Based on the aforementioned analysis, we utilized the ADMETlab 2.0 solver for ML predictions (\href{https://admetmesh.scbdd.com/}{https://admetmesh.scbdd.com/}) \cite{xiong2021admetlab}. Their documentation outlines optimal ranges for various ADMET properties. Additionally, the SAS values were calculated using the Rdkit package \cite{landrum2013rdkit}. Table S2 in the Supporting Information presents the optimal ranges for both ADMET properties and SAS. Notably, a BA value greater than -8.18 kcal/mol is required to minimize hERG side effects. By evaluating ADMET properties, SAS, and cross-target prediction tools, we can systematically identify promising lead compounds.

\begin{figure}[htp]
    \centering
    \includegraphics[width=15cm]{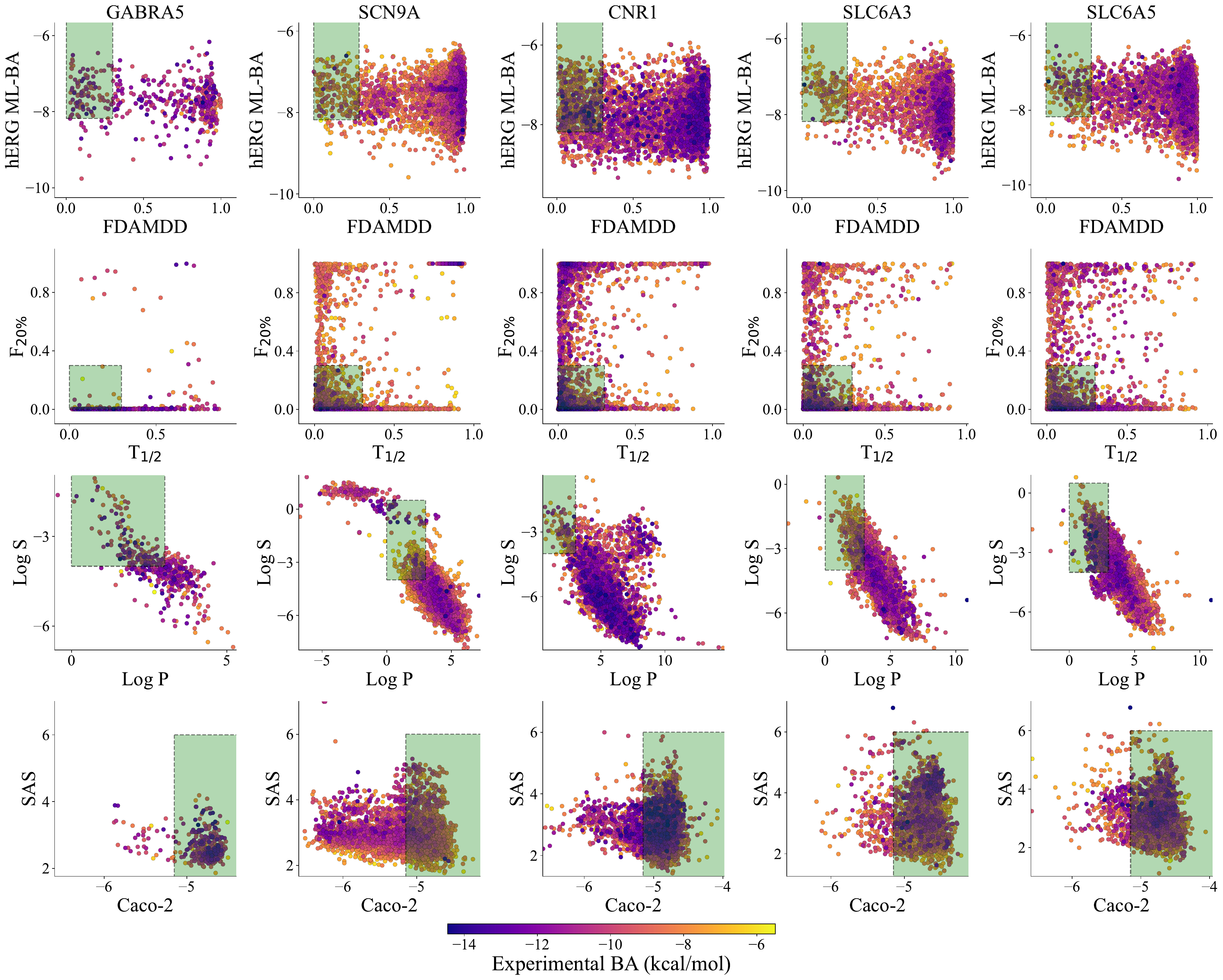}
        \caption{ADMET properties, SAS, and hERG side effects screening for the datasets of GABRA5, SCN9A, CNR1, SLC6A3, and SLC6A5. The color of the scatter points represents the experimental BA values of the compounds in each dataset, with green frames highlighting the optimal range for properties and side effects.}
        \label{fig5}
\end{figure}

Figure \ref{fig5} presents the ADMET screening results for datasets involving GABRA5, SCN9A, CNR1, SLC6A3, and SLC6A5. Specifically, GABRA5 encodes the $\alpha5$ subunit of the GABA\textsubscript{A} receptor, which is targeted by benzodiazepines and barbiturates for sedation. SCN9A encodes the Na\textsubscript{v}1.7 sodium channel, which plays a crucial role in pain perception and anesthetic analgesia. CNR1 encodes the cannabinoid receptor type 1, influencing pain and the effects of cannabinoid anesthetics. SLC6A3 and SLC6A5 encode dopamine and glycine transporters, respectively, which are involved in neurotransmission and anesthetic mechanisms.

In Figure \ref{fig5}, the four rows display the screening results for eight properties across the five datasets, with the color of the scatter points representing the experimental BA values of the compounds. The green frames in each panel delineate the optimal ranges for the pairs of screening properties indicated on the $x$-axis and $y$-axis. Notably, T\textsubscript{1/2} and F\textsubscript{20\%} established stricter screening criteria, with only a small fraction of compounds falling within the green frames. In contrast, the Caco-2 screening encompassed approximately half of the compounds, while SAS screening functioned as a relatively loose filter. Overall, the combined screening of ADMET properties, SAS, and hERG established stringent criteria for identifying nearly optimal lead compounds.

\subsection{Side effect assessment of available anesthetics}
Procaine and tetracaine are both local anesthetics used to manage pain through regional nerve blockade \cite{mertz1990photophysical}. Procaine, also known by its trade name Novocain, is an ester-type local anesthetic that was one of the first agents developed for this purpose. It exerts its anesthetic effects by inhibiting nerve signal transmission; however, its duration of action is relatively short, necessitating the use of adjunct agents to prolong the anesthetic effect. In contrast, tetracaine is a more potent and longer-lasting ester-type local anesthetic compared to procaine. Due to its increased potency and extended duration of action, tetracaine is often employed in procedures requiring prolonged anesthesia, such as ophthalmic surgeries or certain dermatological interventions. While both agents share a similar chemical structure and mechanism of action, their clinical applications are distinguished by differences in efficacy, duration of action, and potential side effects. Our model predicted that procaine and tetracaine have BAs to GABRA5 of -10.71 kcal/mol and -10.50 kcal/mol, respectively, indicating that both medications can effectively bind to GABRA5. Furthermore, we predicted the BA of procaine and tetracaine to hERG to be -7.34 kcal/mol and -8.02 kcal/mol, respectively, suggesting that they are unlikely to cause hERG-related side effects. Moreover, procaine is predicted to be effective against CALCR, NTSR1, APLNR, and SSTR2, with BAs of -12.77, -10.85, -10.63, and -10.61 kcal/mol, respectively. In contrast, tetracaine shows the highest BA for CALCR, S1PR1, NTSR1, and CCKAR, with values of -13.37, -10.81, -10.79, and -10.62 kcal/mol, respectively. Notably, CALCR plays a crucial role in calcium and bone metabolism, and the BA data suggest that procaine and tetracaine may exert certain physiological or pharmacological effects by influencing the function of CALCR. The side effect discussion of more existing anesthetics can be found in the Supporting Information S4.

\subsection{Nearly optimal lead compounds from screening and repurposing}

We aim to discover anesthetics that target GABA receptors through two primary approaches: screening and repurposing. In these processes, we utilize 136 models to predict cross-target  BAs. Beyond addressing potency, it is essential to meet the optimal ranges for ADMET properties, synthetic accessibility (as shown in Table S2 of the Supporting Information), and hERG side effects. GABRA5 is a crucial target for anesthetics, and our goal is to identify promising and effective compounds for this receptor using the 136 datasets as our compound source. For the screening process, we start with potent inhibitors from the GABRA5 dataset, selecting compounds with an experimental BA value of less than -9.54 kcal/mol. We then assess additional properties, ensuring that selected molecules exhibit no side effects on the other 134 protein targets and hERG, which means their BA should be greater than -9.54 kcal/mol. Specifically, for hERG, the BA must exceed -8.18 kcal/mol. In the repurposing process, we evaluate the binding efficacy of all weak inhibitors from the other 135 datasets against GABRA5. We initiate this evaluation with compounds having an experimental BA value greater than -9.54 kcal/mol and identify those predicted to have a BA of less than -9.54 kcal/mol for GABRA5. It is crucial to exclude compounds with side effects on the other 134 protein targets and hERG during this search. Lastly, all compounds must be screened for optimal ADMET properties and synthetic accessibility.

\begin{figure}[htp]
    \centering
    \includegraphics[width=15cm]{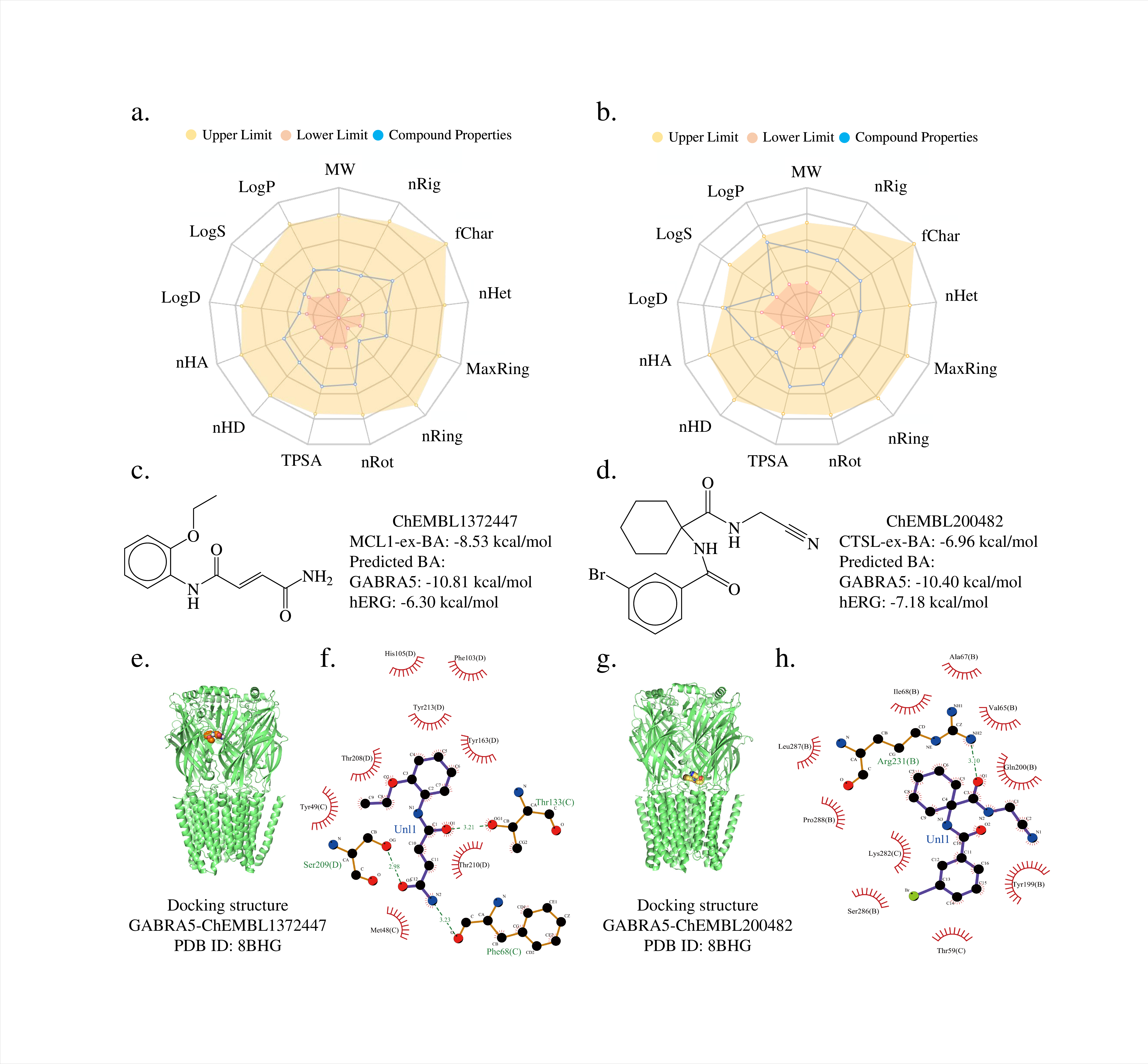}
        \caption{Further evaluation of additional ADMET properties for the identified repurposable molecular compounds is conducted. a and c represent the predicted ADMET properties, chemical graph and side effect assessments for compound ChEMBL1372447, while b and d represent these predictions and graphs for compound ChEMBL200482. In a and b, the boundaries of the yellow and orange regions respectively highlight the upper and lower limits of the optimal range for ADMET properties. The blue curves indicate the values of the specified 13 ADMET properties. The predictions shown in a and b are from the ADMETlab 2.0 website (https://admetmesh.scbdd.com/). The 3D docking structures between compunds ChEMBL1372447, ChEMBL200482, and GABRA5 are shown in e and g. The corresponding 2D interaction diagrams are given in f and h. The PDB ID for the GABRA5 protein is 8BHG. AutoDock Vina was used for protein-ligand docking, and hydrogen bonds play a crucial role in binding energy. Abbreviations: MW (Molecular Weight), $\log$P (log of octanol/water partition coefficient),  $\log$S (log of the aqueous  solubility), $\log$D (logP at physiological pH 7.4), nHA (Number of hydrogen bond acceptors), nHD (Number of hydrogen bond donors), TPSA (Topological polar surface area), nRot (Number of rotatable bonds), nRing (Number of rings), MaxRing (Number of atoms in the biggest ring), nHet (Number of heteroatoms), fChar (Formal charge), and nRig (Number of rigid bonds).}
        \label{fig6}
\end{figure}

Finding compounds that meet all the aforementioned requirements is challenging. Ultimately, we identified two nearly optimal lead compounds: ChEMBL1372447 from the MCL1 dataset and ChEMBL200482 from the CTSL dataset, for repurposing. The BAs of ChEMBL1372447 and ChEMBL200482 for GABRA5 are -10.81 kcal/mol and -10.40 kcal/mol, respectively, demonstrating strong binding efficacy, and thus predicting their effectiveness against GABRA5. Their BA for hERG are -6.30 kcal/mol and -7.18 kcal/mol, respectively, confirming that they do not have side effects on hERG. Additionally, predictions show that these compounds do not exhibit binding activity or side effects on 122 other proteins and 119 other proteins, respectively. Furthermore, we used the ADMETlab 2.0 prediction solver to evaluate more ADMET properties of these two molecular compounds. As illustrated in Figures \ref{fig6}a and \ref{fig6}b, these compounds remain within the optimal range for these ADMET properties. The meanings and optimal ranges of the 13 ADMET properties are provided in Table S8 of the Supporting Information. Figures \ref{fig6} c-d show the chemical graphs and prediction of side effects of these two compounds.

We are also interested in the molecular interactions between these two compounds and the GABRA5 protein structure. To analyze these interactions, we utilized the AutoDock Vina software for protein-ligand docking \cite{huey2012using}. The 3D docking structures and 2D interaction diagrams are shown in Figure \ref{fig6}. The AutoDock Vina software generated nine docking poses, and nine docking scores were calculated based on its scoring function. The details of docking between these two compounds with GABRA5 can be found in Table S9 and S10 of the Supporting Information, respectively. In Figure \ref{fig6}, we have adopted the pose with the highest affinity (kcal/mol). As can be seen from the figure, hydrogen bonds are formed between the compounds and the GABRA5 protein. Specifically, compound ChEMBL1372447 forms three hydrogen bonds with Ser209 (2.98 $\mathring{\rm A}$), Thr133 (3.21 $\mathring{\rm A}$), and Phe68 (3.23 $\mathring{\rm A}$), respectively, while compound ChEMBL200482 forms one hydrogen bond with Arg231 (3.10 $\mathring{\rm A}$). Moreover, neither compound forms covalent bonds with the side chains of the GABRA5 protein, indicating that hydrogen bonds play a significant role in binding energy. Additionally, four medications selected from the DrugBank database were docked with the 3D structure of the GABRA5 protein, and their 2D interaction diagrams were analyzed in Figure S15 of the Supporting Information.


\subsection{Molecular optimization of existing anesthetics}
Molecular optimization of existing anesthetics is of great significance. Through molecular optimization, the efficacy of anesthetics can be enhanced, achieving better anesthetic effects. Additionally, optimization can reduce the side effects caused by anesthetics, thus improving patient safety and comfort. Optimized anesthetics can possess better pharmacokinetic properties. Furthermore, molecular optimization can help develop new anesthetics, reducing the risk of drug resistance and ensuring long-term effectiveness. We collected ADMET properties of some existing anesthetics and found that some anesthetics do not meet the optimal ADMET range. More concerning, some anesthetics exhibit side effects on hERG. Therefore, we decided to optimize these anesthetics using the OptADMET online server (\href{https://cadd.nscc-tj.cn/deploy/optadmet}{https://cadd.nscc-tj.cn/deploy/optadmet/}). OptADMET is the first integrated chemical transformation rules platform that covers 32 key ADMET properties, offering a variety of attribute rules and providing valuable experience for the optimization of lead compounds \cite{yi2024optadmet}.

\begin{figure}[htp]
    \centering
    \includegraphics[width=14cm]{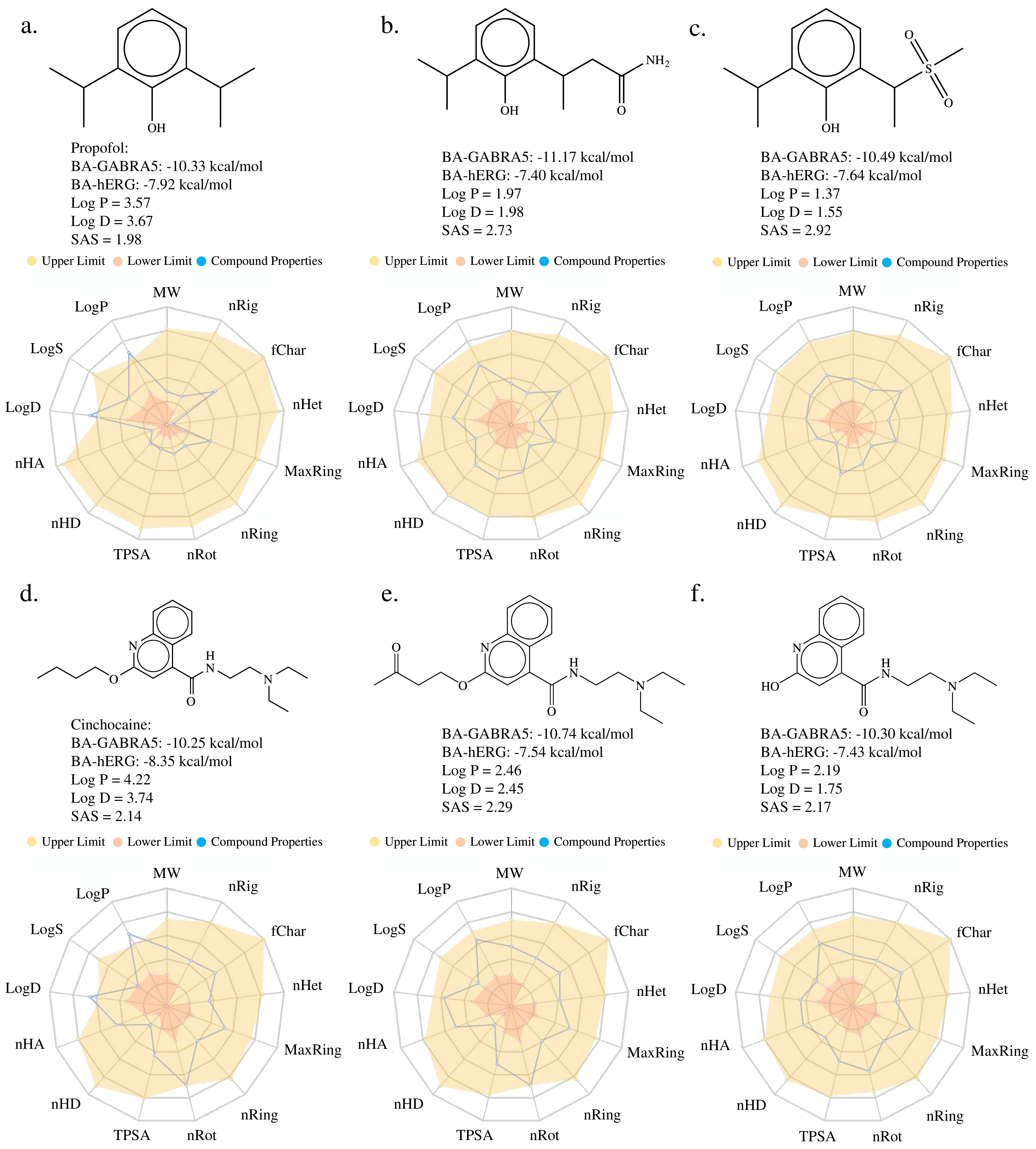}
        \caption{Two anesthetics and their newly optimized molecules, with ADMET profile charts below each molecule. a: Anesthetic propofol and its related properties. b and c: Two optimized molecules of propofol. d: Anesthetic cinchocaine and its related properties. e and f: Two optimized molecules of cinchocaine.}
        \label{fig8}
\end{figure}

As shown in Figure \ref{fig8}, for the anesthetic propofol, its BA to GABRA5 is -10.33 kcal/mol, and to hERG is -7.92 kcal/mol. However, its log P and log D values are 3.57 and 3.67, not within the respective optimal ADMET range 0-3 and 1-3. Therefore, we optimized these properties. The new molecules in Figures \ref{fig8}b and \ref{fig8}c have BAs to GABRA5 of -11.17 kcal/mol and -10.49 kcal/mol, respectively, showing good BA and suitability for anesthesia. Their BAs to hERG are -7.40 and -7.64 kcal/mol, indicating no side effects on hERG. Furthermore, their log P values are 1.97 and 1.37, and log D values are 1.98 and 1.55, respectively, all within the optimal ADMET range. Importantly, their SAS values are 2.73 and 2.92, indicating they are relatively easy to synthesize. For the anesthetic cinchocaine, we found it has side effects on hERG, with a BA of -8.35 kcal/mol. Its log P is 4.22, and log D is 3.74, both exceeding the optimal range. Therefore, we optimized log P, log D, and hERG properties. The new molecules in Figures \ref{fig8}e and \ref{fig8}f have BAs to GABRA5 of -10.74 and -10.30 kcal/mol. More importantly, their BAs to hERG are -7.54 and -7.43 kcal/mol, indicating no side effects on hERG. The optimized log P values are 2.46 and 2.19, and log D values are 2.45 and 1.75, respectively, all within the optimal ADMET range. The SAS values confirm they are easy to synthesize. Hence, through molecular optimization, common anesthetics can achieve better pharmacokinetic properties and reduced side effects.

\subsection{Machine-learning repurposing of DrugBank compounds for anesthetics}
Drug repurposing refers to the research strategy of investigating drugs that have already been marketed, are under development, or whose development has been discontinued, for new therapeutic applications. This approach can reduce the failure rate and cost of drug development, while potentially revealing new targets and therapeutic pathways. In our study, we collected 11,906 drugs from the DrugBank database and predicted their binding affinities for GABRA5 and hERG. Table S11 of the Supporting Information presents the 15 FDA-approved drugs from DrugBank with the highest binding affinities. Our ML model predicts that these compounds are effective against the GABRA5 receptor while exhibiting no adverse effects on hERG. Among these drugs, Chloramphenicol is a broad-spectrum antibiotic known for its effectiveness against various severe bacterial infections. Darolutamide, a non-steroidal androgen receptor antagonist, is utilized in treating non-metastatic, castration-resistant prostate cancer. Maribavir serves as an inhibitor of the cytomegalovirus (CMV) pUL97 kinase, and it is prescribed for treating refractory CMV infections, particularly in post-transplant patients. Upadacitinib, an oral inhibitor that selectively targets Janus kinase (JAK) 1, is used for managing conditions such as moderate to severe rheumatoid arthritis, psoriatic arthritis, ankylosing spondylitis, and severe atopic dermatitis. We predicted the binding affinities of these four drugs to GABRA5 to be -11.94, -11.92, -11.68, and -11.66 kcal/mol, respectively. Hence, these drugs have the potential to be repurposed as anesthetics.

We are also interested in the molecular interactions between these drugs and the GABRA5 protein structure. We used the AutoDock Vina software to perform protein-ligand docking and analyze these interactions. Nine docking poses were generated using AutoDock Vina, and we adopted the pose with the highest affinity (kcal/mol) in our illustrations. Figure S15a, S15c, S15e, and S15g of the Supporting Information are the docking structures of these four drugs bound to GABRA5, respectively, and Table S12, S13, S14, and S15 of the Supporting Information are the details of docking between these four drugs and GABRA5 by AutoDock Vina. For the 2D interaction diagrams of the four drugs, as shown in Figure S15b, Chloramphenicol forms two hydrogen bonds with Thr133 (2.81 $\mathring{\rm A}$) and Ser209 (2.87 $\mathring{\rm A}$), respectively. In Figure S15d, Darolutamide forms two hydrogen bonds with Asp187 (2.91 $\mathring{\rm A}$) and Lys159 (2.80 $\mathring{\rm A}$). Maribavir forms three hydrogen bonds with Phe68 (3.07 $\mathring{\rm A}$), Ser209 (2.67 $\mathring{\rm A}$), and Asp47 (2.99 $\mathring{\rm A}$), along with two hydrogen bonds with Thr133 at distances of 3.12 $\mathring{\rm A}$ and 2.74 $\mathring{\rm A}$, and Upadacitinib forms one hydrogen bond with Ser279 (3.34 $\mathring{\rm A}$) shown in Figure S15f and S15h, respectively. These molecular interactions further validate our predictions. In addition, the remaining 11 drugs listed in Table S11 of the Supporting Information also demonstrate high binding affinities for GABRA5, suggesting their potential for repurposing as anesthetics.

\subsection{Discussion}
Anesthesia and anesthetic agents are crucial in the medical field, alleviating pain during surgical procedures and ensuring successful outcomes. However, existing anesthetics have limitations, including significant side effects, individual variability, and challenges in controlling their effects. Each year, numerous cases of patient mortality arise from anesthesia-related complications globally, and managing anesthetics imposes a substantial economic burden on medical institutions. To enhance safety and reduce risks, pharmaceutical companies and researchers are working to develop new anesthetic agents. Although progress has been made, it remains relatively slow, highlighting the need for more targeted anesthetic agents tailored to different surgeries and patient groups.

Based on protein-protein interaction networks, we developed a corresponding drug-target interaction network and identified two lead compounds for new anesthetic design by proteomic learning. To further validate the anesthetic potential and safety of these compounds, we will carry out a series of animal experiments to determine their agonist or antagonist properties in the future. Concurrently, we will assess the toxicological characteristics and blood-brain barrier permeability of these compounds through \textit{in vitro} experiments and animal models. Our generative network module will also be employed to continuously generate new candidate drugs and predict their side effects, thereby accelerating the development process of novel anesthetic agents \cite{gao2020generative,feng2023multiobjective}. This research not only provides new insights into the innovation of anesthetic agents but also offers more safe and effective treatment options for clinical anesthesia.

Despite the increasing application of machine learning (ML) techniques in anesthetic drug discovery, these methods still face intrinsic challenges and limitations. The complexity of anesthesia itself presents significant obstacles to the advancement of new anesthetics. One major issue is the scarcity of specialized datasets tailored for anesthetic drug research, compounded by the fact that the exact mechanisms underlying anesthesia remain poorly understood. This creates additional barriers to the development of novel drugs. 

It is also important to highlight the rapid advancements in large language models (LLMs), like ChatGPT, which have recently achieved remarkable breakthroughs and drawn significant attention. LLMs have shown substantial promise in bioinformatics and drug discovery tasks, offering new possibilities in these areas. Furthermore, the rapid evolution of proteomics has opened new research pathways in anesthesia by investigating drug mechanisms, identifying personalized treatment biomarkers, and assessing drug safety. 

Looking ahead, integrating LLMs with proteomic technologies could potentially revolutionize the field by enabling the personalization of anesthesia, intensive care, and pain management for individual patients. Such integration could also complement pharmacogenetic approaches to optimizing drug therapies.

\section{Methods}

\subsection{Datasets}

All datasets were collected from the ChEMBL database (\href{https://www.ebi.ac.uk/chembl/}{https://www.ebi.ac.uk/chembl/}), which is used to study proteins within the GABA receptor network \cite{gaulton2017chembl}. The ChEMBL database is a widely used repository of bioactive compound information, developed and maintained by the European Bioinformatics Institute (EMBL-EBI). It provides a wealth of data on drug chemistry and bioactivity, including compound structures, targets, mechanisms of action, pharmacokinetic properties, and toxicity information. Since ML training requires a sufficient number of data points, we set the minimum data size at 250, ultimately resulting in 136 datasets. The metrics for these datasets are IC$_{50}$ and $\rm K_{i}$. IC$_{50}$ (half-maximal inhibitory concentration) refers to the concentration of a compound needed to inhibit a specific biological process or enzyme activity by half. $\rm K_{i}$ (inhibition constant) measures the binding strength of an inhibitor to an enzyme or receptor, with a lower $\rm K_{i}$ indicating a higher BA. IC$_{50}$ can be approximately converted to $\rm K_{i}$ using the formula K$_i$=IC$_{50}/2$ \cite{kalliokoski2013comparability}. Subsequently, using these metrics, we calculate BA with the formula BA=1.3633$\rm \times \log_{10} K_{i}$ (kcal/mol) to build ML models. Additionally, since blocking the hERG channel can lead to fatal arrhythmias, it is crucial to avoid hERG-related side effects. Therefore, we also collected datasets of hERG inhibitors. Detailed information about all datasets can be found in the Supporting Information.

\subsection{Molecular embeddings}
In this study, we collected 136 datasets where molecular representations are provided as 2D SMILES strings. SMILES, using ASCII strings, precisely depict molecular structures including atoms, bonds, and cyclic formations, simplifying input and representation of complex molecules. Consequently, two forms of molecular fingerprints were generated using pre-trained models based on natural language processing (NLP) algorithms, specifically bidirectional encoder transformers \cite{chen2021extracting,chen2021algebraic} and sequence-to-sequence autoencoders \cite{winter2019learning}. These models encode the 2D SMILES strings of compounds into latent embedding vectors of length 512, thus producing 512-dimensional features. We denote the fingerprints derived from the transformer and autoencoder models as BET-FP and AE-FP, respectively.

\subsubsection{Bidirectional encoder transformer}
In recent years, Chen et al. developed a self-supervised learning (SSL) platform to pre-train deep learning networks on millions of unlabeled molecular data \cite{chen2021extracting,chen2021algebraic}. The platform utilizes the Bidirectional Encoder Transformer (BET) model, leveraging attention mechanisms for effective molecular representation learning. Unlike traditional encoder-decoder frameworks, the SSL platform solely employs the encoder network to encode SMILES strings, thereby simplifying the model architecture \cite{jiang2024transformer,jiang2024review}.

Prior to the commencement of training, the SMILES strings undergo preprocessing. Each symbol in the SMILES string is treated as a constituent part, amounting to a total of 51 symbols. The SMILES strings serve as the model's input, with a maximum length requirement of 256. To facilitate model processing, special symbols $'\langle s \rangle'$ and $'\langle \backslash s \rangle'$ are appended at the beginning and end of the SMILES strings, respectively. If the length of a SMILES string is less than 256, the $'\langle pad \rangle'$ symbol is used for padding.

To enable self-supervised learning, data masking is applied to the SMILES strings. A random subset of 15\% of the symbols in all SMILES strings is selected for manipulation, with 80\% being masked, 10\% remaining unchanged, and the remaining 10\% being randomly replaced. This masking strategy effectively simulates unlabeled data and guides the model in learning molecular representations.

The BET model consists of eight bidirectional encoder layers, each containing a multi-head self-attention layer followed by a fully connected feed-forward neural network. The self-attention layer comprises 8 heads, each with an embedding dimension of 512. The Adam optimizer is utilized for model training, and a weight decay strategy is applied. The loss function is defined as cross-entropy, which measures the discrepancy between the model’s predicted values and the true values.

The average embedding vector of all valid symbols within a SMILES string is computed, resulting in a final molecular embedding matrix. Each SMILES string corresponds to a 256-dimensional embedding vector, representing the molecule’s fingerprint. To evaluate the performance of the pre-trained model on various datasets, SMILES strings from the ChEMBL database are employed for pre-training, yielding a pre-trained model. The experimental results demonstrate that this model effectively generates molecular fingerprints with strong predictive capabilities and is suitable for downstream tasks.

\subsubsection{Sequence-to-sequence auto-encoder}
Winter and colleagues have developed a novel unsupervised deep learning approach aimed at extracting implicit chemical information from SMILES molecular structure representations \cite{winter2019learning}. The core of this method is a sequence-to-sequence autoencoder that is capable of translating one molecular representation form into another, compressing the complete description of the chemical structure into the latent representation between the encoder and decoder. When translated into another semantically equivalent but syntactically distinct molecular representation, the model extracts physicochemical information embedded within the molecular representation. This model is trained on a large-scale dataset of chemical structures, allowing for the extraction of molecular descriptors for query structures without the need for retraining or incorporating labels.

The translation model consists of an encoder and decoder network. The encoder takes a molecular structure representation (e.g., SMILES) as input and encodes it into a low-dimensional continuous vector representation, known as the latent space representation. The encoder can employ either convolutional neural network (CNN) or recurrent neural network (RNN) architectures, followed by a fully connected layer that maps the output to the latent space. The decoder, on the other hand, takes the latent space representation as input and generates another molecular structure representation (e.g., InChI). The decoder utilizes an RNN architecture, with its cell states initialized by an individual fully connected layer for each layer in the RNN. To further encourage the model to learn more meaningful molecular representations, an additional classification model can be incorporated. This model takes the latent space representation as input and predicts molecular properties that can be directly inferred from the molecular structure.

The translation model was trained on approximately 72 million molecular compounds from the ZINC and PubChem databases. The compounds underwent preprocessing to filter out those that meet various criteria, including molecular weight, number of heavy atoms, partition coefficient, and other properties. After extensive training on the preprocessed dataset, the resulting translation model generates embedding vectors that serve as molecular fingerprints.

\subsection{Machine-learning models}
In constructing ML models, we have employed three ML algorithms, namely Gradient Boosting Decision Tree (GBDT), Support Vector Machine (SVM), and Random Forest (RF) \cite{gao2021proteome,feng2022machine,feng2023machine2}. GBDT is an algorithm based on ensemble learning technology that enhances the predictive performance of the model by constructing multiple decision trees and combining their predictive results. The core concept is to use the gradient descent algorithm to optimize the loss function, thereby progressively learning the weights of each decision tree. The GBDT model is capable of capturing the nonlinear relationships within the data and exhibits high generalization capability.

SVM, on the other hand, is a classical supervised learning algorithm that classifies data by finding the hyperplane with the maximum margin. The SVM model is effective in handling high-dimensional data and possesses strong generalization ability. Moreover, SVM can address nonlinear problems through the use of kernel functions, enabling it to manage more complex datasets.

RF, similar to GBDT, is an algorithm rooted in ensemble learning technology that boosts the model’s predictive performance by constructing multiple decision trees and integrating their predictions. However, unlike GBDT, RF randomly selects a subset of features and samples when building each decision tree. This approach allows RF to better capture the intricate relationships within the data and reduces the risk of overfitting. The RF model is known for its robustness and generalization ability, and it is widely applied across various classification and regression tasks.

In total, we have gathered 136 datasets, each containing no fewer than 250 data points. As detailed in Table S4  of the Supporting Information, following a comparative analysis of the three models, it is recommended to utilize the SVM algorithm for constructing models for these datasets. As previously mentioned, two types of molecular fingerprints BET-FP and AE-FP were employed to represent the compounds. Our ML models were developed by integrating these molecular fingerprints with the GBDT algorithm. We have established a total of 136 ligand-based ML models for the 136 datasets. For each dataset, two separate models were constructed by pairing BET-FP and AE-FP with the SVM algorithm, respectively, and the average prediction of the two individual models was taken as our final BA prediction. Typically, this averaged or consensus approach yields predictions that outperform those of individual models. To mitigate the impact of randomness, each individual SVM model was trained ten times with different random seeds. The average of the ten predictions was adopted as the final result for each model. In Table S1  of the Supporting Information, we include the Pearson correlation coefficients from ten fold cross-validation used for modeling the 136 datasets.

\section{Conclusion}
The GABA receptor serves as a pivotal target for anesthetic agents, playing a crucial role in modulating neurotransmitter balance and inducing anesthetic effects. The development of new anesthetic agents necessitates a deep understanding of the complex interactions between drugs and the GABA receptor. In this study, we introduce a proteomic learning strategy to explore potential candidates for novel anesthetic agents based on 24 anesthesia related GABA receptors. We consider 4824 targets within 24 protein-protein interaction (PPI) networks and 1504529 known binding compounds from ChEMBL database, and curate 980 unique targets. Through the data selection conditions that the inhibitor compound should be homo sapiens and single protein, as well as the minimal training number 250 to ensure the reliable prediction,  we construct the corresponding drug-target network (DTI) and ultimately identify 136 targets as side effect targets with a total of 183250 inhibitor compounds to screen potential lead compounds for novel anesthetic design. 

In the realm of anesthesiology, our ML platform offers a new strategy for discovering potential anesthetic agents. This innovative approach has the capacity to be broadly applied to research on various conditions that impact the nervous system. With the continuous refinement of our knowledge about the mechanisms of anesthesia and the ongoing quest for safer and more effective anesthetic treatments, our platform is poised to tackle the significant challenges faced in the field of anesthesia. By leveraging this technology, we can not only enhance the development of new anesthetic compounds but also contribute to the improvement of patient outcomes and the reduction of anesthesia-related risks, thereby addressing critical public health concerns associated with anesthesia and perioperative care.

\section*{Data and code availability}
The related codes studied in present work are available at: https://github.com/LongChen0/GABA-PPI

\section*{Acknowledgements}
This work was supported in part by NIH grants R01GM126189 and R01AI164266, NSF grants DMS-2052983, DMS-1761320, and IIS-1900473, NASA grant 80NSSC21M0023, MSU Foundation, Bristol-Myers Squibb 65109, and Pfizer. The work of Huahai Qiu and Bengong Zhang was supported by the National Natural Science Foundation of China under Grant No.12271416 and No.12371500, respectively.

\section*{Competing interests}
The authors declare no competing interests.


\begin{thebibliography}{10}

\bibitem{abdelgadir2024determine}
E.~H. Abdelgadir, S.~D. Alshehri, and S.~Kumar.
\newblock Determine the pharmacokinetics (half-life, volume of distribution and clearance) of amb-fubinaca in rats plasma using gc--ms/ms.
\newblock {\em Journal of Pharmacological and Toxicological Methods}, 127:107513, 2024.

\bibitem{ballester2010machine}
P.~J. Ballester and J.~B. Mitchell.
\newblock A machine learning approach to predicting protein--ligand binding affinity with applications to molecular docking.
\newblock {\em Bioinformatics}, 26(9):1169--1175, 2010.

\bibitem{bodor1989new}
N.~Bodor, Z.~Gabanyi, and C.~K. Wong.
\newblock A new method for the estimation of partition coefficient.
\newblock {\em Journal of the American Chemical Society}, 111(11):3783--3786, 1989.

\bibitem{chen2021algebraic}
D.~Chen, K.~Gao, D.~D. Nguyen, X.~Chen, Y.~Jiang, G.-W. Wei, and F.~Pan.
\newblock Algebraic graph-assisted bidirectional transformers for molecular property prediction.
\newblock {\em Nature communications}, 12(1):3521, 2021.

\bibitem{chen2021extracting}
D.~Chen, J.~Zheng, G.-W. Wei, and F.~Pan.
\newblock Extracting predictive representations from hundreds of millions of molecules.
\newblock {\em The journal of physical chemistry letters}, 12(44):10793--10801, 2021.

\bibitem{chen2020mutations}
J.~Chen, R.~Wang, M.~Wang, and G.-W. Wei.
\newblock Mutations strengthened sars-cov-2 infectivity.
\newblock {\em Journal of molecular biology}, 432(19):5212--5226, 2020.

\bibitem{chen2024machine}
L.~Chen, J.~Jiang, B.~Dou, H.~Feng, J.~Liu, Y.~Zhu, B.~Zhang, T.~Zhou, and G.-W. Wei.
\newblock Machine learning study of the extended drug--target interaction network informed by pain related voltage-gated sodium channels.
\newblock {\em Pain}, 165(4):908--921, 2024.

\bibitem{crowder2024systematized}
C.~M. Crowder and S.~A. Forman.
\newblock Systematized serendipity: Fishing expeditions for anesthetic drugs and targets.
\newblock {\em Anesthesiology}, pages 10--1097, 2024.

\bibitem{du2024multiscale}
H.~Du, G.-W. Wei, and T.~Hou.
\newblock Multiscale topology in interactomic network: from transcriptome to antiaddiction drug repurposing.
\newblock {\em Briefings in Bioinformatics}, 25(2):bbae054, 2024.

\bibitem{feng2023machine2}
H.~Feng, R.~Elladki, J.~Jiang, and G.-W. Wei.
\newblock Machine-learning analysis of opioid use disorder informed by mor, dor, kor, nor and zor-based interactome networks.
\newblock {\em Computers in biology and medicine}, 157:106745, 2023.

\bibitem{feng2022machine}
H.~Feng, K.~Gao, D.~Chen, L.~Shen, A.~J. Robison, E.~Ellsworth, and G.-W. Wei.
\newblock Machine learning analysis of cocaine addiction informed by dat, sert, and net-based interactome networks.
\newblock {\em Journal of chemical theory and computation}, 18(4):2703--2719, 2022.

\bibitem{feng2023machine}
H.~Feng, J.~Jiang, and G.-W. Wei.
\newblock Machine-learning repurposing of drugbank compounds for opioid use disorder.
\newblock {\em Computers in biology and medicine}, 160:106921, 2023.

\bibitem{feng2023multiobjective}
H.~Feng, R.~Wang, C.-G. Zhan, and G.-W. Wei.
\newblock Multiobjective molecular optimization for opioid use disorder treatment using generative network complex.
\newblock {\em Journal of medicinal chemistry}, 66(17):12479--12498, 2023.

\bibitem{feng2023virtual}
H.~Feng and G.-W. Wei.
\newblock Virtual screening of drugbank database for herg blockers using topological laplacian-assisted ai models.
\newblock {\em Computers in biology and medicine}, 153:106491, 2023.

\bibitem{flower2002drug}
D.~R. Flower.
\newblock {\em Drug design: cutting edge approaches}, volume 279.
\newblock Royal Society of Chemistry, 2002.

\bibitem{fritzius2024preassembly}
T.~Fritzius, R.~Ture{\v{c}}ek, D.~Fernandez-Fernandez, S.~Isogai, P.~D. Rem, M.~Kralikova, M.~Gassmann, and B.~Bettler.
\newblock Preassembly of specific g$\beta$$\gamma$ subunits at gabab receptors through auxiliary kctd proteins accelerates channel gating.
\newblock {\em Biochemical Pharmacology}, page 116176, 2024.

\bibitem{gao2021proteome}
K.~Gao, D.~Chen, A.~J. Robison, and G.-W. Wei.
\newblock Proteome-informed machine learning studies of cocaine addiction.
\newblock {\em The journal of physical chemistry letters}, 12(45):11122--11134, 2021.

\bibitem{gao2020generative}
K.~Gao, D.~D. Nguyen, M.~Tu, and G.-W. Wei.
\newblock Generative network complex for the automated generation of drug-like molecules.
\newblock {\em Journal of chemical information and modeling}, 60(12):5682--5698, 2020.

\bibitem{gaulton2017chembl}
A.~Gaulton, A.~Hersey, M.~Nowotka, A.~P. Bento, J.~Chambers, D.~Mendez, P.~Mutowo, F.~Atkinson, L.~J. Bellis, E.~Cibri{\'a}n-Uhalte, et~al.
\newblock The chembl database in 2017.
\newblock {\em Nucleic acids research}, 45(D1):D945--D954, 2017.

\bibitem{hassan2024sclareol}
S.~H. Hassan, H.~A. El-Nashar, M.~A. Rahman, J.~I. Polash, M.~H. Bappi, M.~Mondal, M.~A. Abdel-Maksoud, A.~Malik, M.~Aufy, M.~El-Shazly, et~al.
\newblock Sclareol antagonizes the sedative effect of diazepam in thiopental sodium-induced sleeping animals: In vivo and in silico studies.
\newblock {\em Biomedicine \& Pharmacotherapy}, 176:116939, 2024.

\bibitem{huey2012using}
R.~Huey, G.~M. Morris, S.~Forli, et~al.
\newblock Using autodock 4 and autodock vina with autodocktools: a tutorial.
\newblock {\em The Scripps Research Institute Molecular Graphics Laboratory}, 10550(92037):1000, 2012.

\bibitem{jiang2024review}
J.~Jiang, L.~Chen, L.~Ke, B.~Dou, C.~Zhang, H.~Feng, Y.~Zhu, H.~Qiu, B.~Zhang, and G.~Wei.
\newblock A review of transformers in drug discovery and beyond.
\newblock {\em Journal of Pharmaceutical Analysis}, page 101081, 2024.

\bibitem{jiang2024transformer}
J.~Jiang, L.~Ke, L.~Chen, B.~Dou, Y.~Zhu, J.~Liu, B.~Zhang, T.~Zhou, and G.-W. Wei.
\newblock Transformer technology in molecular science.
\newblock {\em Wiley Interdisciplinary Reviews: Computational Molecular Science}, 14(4):e1725, 2024.

\bibitem{kalliokoski2013comparability}
T.~Kalliokoski, C.~Kramer, A.~Vulpetti, and P.~Gedeck.
\newblock Comparability of mixed ic50 data--a statistical analysis.
\newblock {\em PloS one}, 8(4):e61007, 2013.

\bibitem{karlsson2024ligand}
E.~Karlsson, O.~And{\'e}n, C.~Fan, Z.~Fourati, A.~Haouz, Y.~Zhuang, M.~H. Delarue, R.~J. Howard, and E.~R. Lindahl.
\newblock Ligand-gated ion channel modulation by stimulant derivatives via a vestibular cavity.
\newblock {\em Biophysical Journal}, 123(3):160a, 2024.

\bibitem{karunarathne2023gamma}
W.~A. H.~M. Karunarathne, Y.~H. Choi, M.-H. Lee, C.-H. Kang, and G.-Y. Kim.
\newblock Gamma-aminobutyric acid (gaba)-mediated bone formation and its implications for anti-osteoporosis strategies: Exploring the relation between gaba and gaba receptors.
\newblock {\em Biochemical Pharmacology}, 218:115888, 2023.

\bibitem{koh2023gaba}
W.~Koh, H.~Kwak, E.~Cheong, and C.~J. Lee.
\newblock Gaba tone regulation and its cognitive functions in the brain.
\newblock {\em Nature Reviews Neuroscience}, 24(9):523--539, 2023.

\bibitem{kotekar2018postoperative}
N.~Kotekar, A.~Shenkar, and R.~Nagaraj.
\newblock Postoperative cognitive dysfunction--current preventive strategies.
\newblock {\em Clinical interventions in aging}, pages 2267--2273, 2018.

\bibitem{landrum2013rdkit}
G.~Landrum et~al.
\newblock Rdkit: A software suite for cheminformatics, computational chemistry, and predictive modeling.
\newblock {\em Greg Landrum}, 8(31.10):5281, 2013.

\bibitem{li2024crossfuse}
Q.~Li, Y.~He, and J.~Pan.
\newblock Crossfuse-xgboost: accurate prediction of the maximum recommended daily dose through multi-feature fusion, cross-validation screening and extreme gradient boosting.
\newblock {\em Briefings in Bioinformatics}, 25(1):bbad511, 2024.

\bibitem{lopes2024artificial}
S.~Lopes, G.~Rocha, and L.~Guimar{\~a}es-Pereira.
\newblock Artificial intelligence and its clinical application in anesthesiology: a systematic review.
\newblock {\em Journal of Clinical Monitoring and Computing}, 38(2):247--259, 2024.

\bibitem{m2013pharmacokinetic}
K.~M~Honorio, T.~L~Moda, and A.~D~Andricopulo.
\newblock Pharmacokinetic properties and in silico adme modeling in drug discovery.
\newblock {\em Medicinal Chemistry}, 9(2):163--176, 2013.

\bibitem{mertz1990photophysical}
C.~Mertz, A.~Marques, L.~N. Williamson, and C.~Lin.
\newblock Photophysical studies of local anesthetics, tetracaine and procaine: drug aggregations.
\newblock {\em Photochemistry and photobiology}, 51(4):427--437, 1990.

\bibitem{muheyati2024extrasynaptic}
A.~Muheyati, S.~Jiang, N.~Wang, G.~Yu, and R.~Su.
\newblock Extrasynaptic gabaa receptors in central medial thalamus mediate anesthesia in rats.
\newblock {\em European Journal of Pharmacology}, 972:176561, 2024.

\bibitem{nguyen2019mathematical}
D.~D. Nguyen, Z.~Cang, K.~Wu, M.~Wang, Y.~Cao, and G.-W. Wei.
\newblock Mathematical deep learning for pose and binding affinity prediction and ranking in d3r grand challenges.
\newblock {\em Journal of computer-aided molecular design}, 33:71--82, 2019.

\bibitem{nguyen2020mathdl}
D.~D. Nguyen, K.~Gao, M.~Wang, and G.-W. Wei.
\newblock Mathdl: mathematical deep learning for d3r grand challenge 4.
\newblock {\em Journal of computer-aided molecular design}, 34:131--147, 2020.

\bibitem{qian2023current}
X.~Qian, X.~Zhao, L.~Yu, Y.~Yin, X.-D. Zhang, L.~Wang, J.-X. Li, Q.~Zhu, and J.-L. Luo.
\newblock Current status of gaba receptor subtypes in analgesia.
\newblock {\em Biomedicine \& Pharmacotherapy}, 168:115800, 2023.

\bibitem{saravanan2024deep}
K.~M. Saravanan, J.-F. Wan, L.~Dai, J.~Zhang, J.~Z. Zhang, and H.~Zhang.
\newblock A deep learning based multi-model approach for predicting drug-like chemical compound’s toxicity.
\newblock {\em Methods}, 226:164--175, 2024.

\bibitem{shan2017mirnas}
L.~Shan, D.~Ma, C.~Zhang, W.~Xiong, and Y.~Zhang.
\newblock mirnas may regulate gabaergic transmission associated genes in aged rats with anesthetics-induced recognition and working memory dysfunction.
\newblock {\em Brain research}, 1670:191--200, 2017.

\bibitem{shen2023svsbi}
L.~Shen, H.~Feng, Y.~Qiu, and G.-W. Wei.
\newblock Svsbi: sequence-based virtual screening of biomolecular interactions.
\newblock {\em Communications biology}, 6(1):536, 2023.

\bibitem{song2023necessity}
B.~Song, M.~Zhou, and J.~Zhu.
\newblock Necessity and importance of developing ai in anesthesia from the perspective of clinical safety and information security.
\newblock {\em Medical Science Monitor: International Medical Journal of Experimental and Clinical Research}, 29:e938835--1, 2023.

\bibitem{sun2008caco}
H.~Sun, E.~C. Chow, S.~Liu, Y.~Du, and K.~S. Pang.
\newblock The caco-2 cell monolayer: usefulness and limitations.
\newblock {\em Expert opinion on drug metabolism \& toxicology}, 4(4):395--411, 2008.

\bibitem{swain2017adjuvants}
A.~Swain, D.~S. Nag, S.~Sahu, and D.~P. Samaddar.
\newblock Adjuvants to local anesthetics: Current understanding and future trends.
\newblock {\em World journal of clinical cases}, 5(8):307, 2017.

\bibitem{szklarczyk2019string}
D.~Szklarczyk, A.~L. Gable, D.~Lyon, A.~Junge, S.~Wyder, J.~Huerta-Cepas, M.~Simonovic, N.~T. Doncheva, J.~H. Morris, P.~Bork, et~al.
\newblock String v11: protein--protein association networks with increased coverage, supporting functional discovery in genome-wide experimental datasets.
\newblock {\em Nucleic acids research}, 47(D1):D607--D613, 2019.

\bibitem{takano2024stereochemical}
R.~Takano, R.~Tanaka, K.~Nakamura, H.~Tabata, T.~Oshitari, H.~Natsugari, and H.~Takahashi.
\newblock Stereochemical properties of quazepam and its affinity for the gabaa receptor.
\newblock {\em Bioorganic \& Medicinal Chemistry Letters}, 110:129854, 2024.

\bibitem{tian2011adme}
S.~Tian, Y.~Li, J.~Wang, J.~Zhang, and T.~Hou.
\newblock Adme evaluation in drug discovery. 9. prediction of oral bioavailability in humans based on molecular properties and structural fingerprints.
\newblock {\em Molecular pharmaceutics}, 8(3):841--851, 2011.

\bibitem{vlisides2012neurotoxicity}
P.~Vlisides and Z.~Xie.
\newblock Neurotoxicity of general anesthetics: an update.
\newblock {\em Current Pharmaceutical Design}, 18(38):6232--6240, 2012.

\bibitem{wang2021mechanisms}
R.~Wang, J.~Chen, and G.-W. Wei.
\newblock Mechanisms of sars-cov-2 evolution revealing vaccine-resistant mutations in europe and america.
\newblock {\em The journal of physical chemistry letters}, 12(49):11850--11857, 2021.

\bibitem{win2023using}
Z.-M. Win, A.~M. Cheong, and W.~S. Hopkins.
\newblock Using machine learning to predict partition coefficient (log p) and distribution coefficient (log d) with molecular descriptors and liquid chromatography retention time.
\newblock {\em Journal of Chemical Information and Modeling}, 63(7):1906--1913, 2023.

\bibitem{winter2019learning}
R.~Winter, F.~Montanari, F.~No{\'e}, and D.-A. Clevert.
\newblock Learning continuous and data-driven molecular descriptors by translating equivalent chemical representations.
\newblock {\em Chemical science}, 10(6):1692--1701, 2019.

\bibitem{xiong2021admetlab}
G.~Xiong, Z.~Wu, J.~Yi, L.~Fu, Z.~Yang, C.~Hsieh, M.~Yin, X.~Zeng, C.~Wu, A.~Lu, et~al.
\newblock Admetlab 2.0: an integrated online platform for accurate and comprehensive predictions of admet properties.
\newblock {\em Nucleic acids research}, 49(W1):W5--W14, 2021.

\bibitem{yang2024comparison}
Y.~Yang, Z.~Lang, X.~Wang, P.~Yang, N.~Meng, Y.~Xing, and Y.~Liu.
\newblock Comparison of the efficacy and safety of ciprofol and propofol in sedating patients in the operating room and outside the operating room: a meta-analysis and systematic review.
\newblock {\em BMC anesthesiology}, 24(1):218, 2024.

\bibitem{yi2024optadmet}
J.~Yi, S.~Shi, L.~Fu, Z.~Yang, P.~Nie, A.~Lu, C.~Wu, Y.~Deng, C.~Hsieh, X.~Zeng, et~al.
\newblock Optadmet: a web-based tool for substructure modifications to improve admet properties of lead compounds.
\newblock {\em Nature Protocols}, pages 1--17, 2024.

\bibitem{zeng2024personalized}
S.~Zeng, Q.~Qing, W.~Xu, S.~Yu, M.~Zheng, H.~Tan, J.~Peng, and J.~Huang.
\newblock Personalized anesthesia and precision medicine: a comprehensive review of genetic factors, artificial intelligence, and patient-specific factors.
\newblock {\em Frontiers in Medicine}, 11:1365524, 2024.

\bibitem{zhang2023emerging}
S.~Zhang, Y.~Wang, S.~Zhang, C.~Huang, Q.~Ding, J.~Xia, D.~Wu, and W.~Gao.
\newblock Emerging anesthetic nanomedicines: Current state and challenges.
\newblock {\em International Journal of Nanomedicine}, pages 3913--3935, 2023.

\bibitem{zhu2023tidal}
Z.~Zhu, B.~Dou, Y.~Cao, J.~Jiang, Y.~Zhu, D.~Chen, H.~Feng, J.~Liu, B.~Zhang, T.~Zhou, et~al.
\newblock Tidal: topology-inferred drug addiction learning.
\newblock {\em Journal of chemical information and modeling}, 63(5):1472--1489, 2023.

\bibitem{zurek2012inhibition}
A.~A. Zurek, E.~M. Bridgwater, and B.~A. Orser.
\newblock Inhibition of $\alpha$5 $\gamma$-aminobutyric acid type a receptors restores recognition memory after general anesthesia.
\newblock {\em Anesthesia and Analgesia}, 114(4):845--855, 2012.

\end{thebibliography}
\end{document}